# Effect of the COVID-19 pandemic on bike-sharing demand and hire time: Evidence from Santander Cycles in London

Shahram Heydari[1*], Garyfallos Konstantinoudis[2], Abdul Wahid Behsoodi[1]

[1] Transportation Research Group, Department of Civil, Maritime and Environmental Engineering,
University of Southampton, Southampton, UK
[2] MRC Centre for Environment and Health, Department of Epidemiology and Biostatistics, School of Public Health,
Imperial College London, London, UK

## Abstract

The COVID-19 pandemic has been influencing travel behaviour in many urban areas around the world since the beginning of 2020. As a consequence, bike-sharing schemes have been affected — partly due to the change in travel demand and behaviour as well as a shift from public transit. This study estimates the varying effect of the COVID-19 pandemic on the London bike-sharing system (Santander Cycles) over the period March-December 2020. We employed a Bayesian second-order random walk time-series model to account for temporal correlation in the data. We compared the observed number of cycle hires and hire time with their respective counterfactuals (what would have been if the pandemic had not happened) to estimate the magnitude of the change caused by the pandemic. The results indicated that following a reduction in cycle hires in March and April 2020, the demand rebounded from May 2020, remaining in the expected range of what would have been if the pandemic had not occurred. This could indicate the resiliency of Santander Cycles. With respect to hire time, an important increase occurred in April, May, and June 2020, indicating that bikes were hired for longer trips, perhaps partly due to a shift from public transit.

## 1. Introduction

Cycling as a sustainable mode of travel is proven to be associated with several benefits such as reducing motorised traffic in urban areas, reducing greenhouse gas emissions, reducing the need for parking spaces, and improving mental and physical health due to an increase in physical activity [1–6]. In recent years, several interventions and policy instruments have been devised to reduce the negative externalities of motorised traffic and encourage more people to cycle [5–7]. Bicycle-sharing is one of the key policy interventions integrated into many urban transportation networks across the world, with the aim of promoting cycling [7, 8]. In the past two decades, bike-sharing systems, which allow for flexible and environmentally-friendly travel, have been gaining popularity in many cities around the world.

The London bike-sharing scheme (Santander Cycles), one of the largest schemes in Europe, was launched in Central London on 30 July 2010 with 5000 bikes available at 315 docking stations located in 8 boroughs [9]. Over the last decade, the scheme has rapidly expanded to other areas of the city, covering around 100 square kilometres [9]. As of January 2020, the London bike-sharing system had more than 12,000 bicycles [9]. In terms of travel demand, more than 87 million cycle hires were made within 10 years from its launch, and there were more than 1.7 million cycle hires in 2019 [9].

Due to the COVID-19 pandemic, many countries introduced stringent measures to contain the spread of the virus. This resulted in important changes in travel behaviour [10], especially in urban settings;

---

[*] Corresponding author; email: s.heydari@soton.ac.uk

for example, as a consequence of restrictions on travelling, and safety concerns regarding travelling on public transit [11–13], and reduction in the frequency of public transport services. In this regard, for example, Noland [14], discusses the relationship between mobility and the reproduction rate of COVID-19. Also, previous studies showed that public transport disruptions either internally (e.g., maintenance and strikes) or externally (e.g., natural disasters and outbreak of communicable diseases) have spillover effects on the demand for bike-sharing schemes [15–17]. In fact, such disruptions often cause a temporary increase in the demand for other modes of transport, including bike-sharing schemes. This said, we would expect important changes in the London bike-sharing system due to the recent pandemic.

## 1.1 Literature review

Over the last decade, several studies investigated various aspects of bike-sharing schemes [2, 5, 7, 8, 18–23]. These studies mostly focus on understanding the demand for bike-sharing systems by revealing the impact of built environment, sociodemographic, weather conditions, and policies on bike-sharing services. In fact, multiple factors such as weather conditions and disruptions in public transport services affect the number and duration of cycle hires [7, 8, 12, 16, 17, 24–26]. Evidence shows that weather conditions play a key role not only in explaining the demand for bike-sharing systems, but also in explaining the duration of cycle hires. Previous studies suggest that, in general, there is a positive relationship between temperature and the number of cycle hires [7, 27, 28]. For example, Morton [29] and Chibwe et al. [8], found that the demand for bike-sharing systems increases as temperature increases. However, research shows that temperatures exceeding 30 C reduce the demand for bike-sharing systems in some regions [7, 30]. With respect to trip duration, Gebhart and Noland [24], investigating the impact of weather on bike-share trips in Washington D.C., found that shorter trip durations occur at lower temperatures between 10 F and 49 F, compared to higher temperatures between 50 F to 59 F. Faghih-Imani and Eluru [26], examining Citi Bike in New York, found that trip duration for non-member users are longer than the member users in favourable weather conditions.

With respect to the demand for bike-sharing schemes, several previous studies suggest that hire numbers decrease as rainfall, wind speed, and humidity increase [7, 24, 28–30]. For example, El-Assi et al. [28], found that rainfall and high humidity are unfavourable weather conditions for the Toronto bike-sharing system. Similarly, Morton [29] found a negative correlation between the demand for the London bike-sharing system and rainfall, wind speed, and humidity. Gebhart and Noland [24] found that wind speed and humidity had a negative impact on the demand for the Capital bikeshare scheme. Chibwe et al. [8] found that rainfall, wind speed, and humidity were negatively associated with the demand for the London bike-sharing system.

As discussed by Wang and Noland [12], bike-sharing schemes help improve the resilience of urban transportation networks since they serve as a substitute for public transport services when these are disrupted. This is in accordance with previous research that shows public transport disruptions (including safety concerns) shift the demand from public transit to bike-sharing systems [11, 16, 31]. Therefore, not only the change in travel behaviour due to the COVID-19 pandemic affects bike-sharing systems, but also changes in public transport (e.g., lower frequency and safety concerns relating to the danger of contracting the virus) have an impact on bike-sharing.

For example, Saberi et al. [16] examining the impacts of a London Tube strike on the London bike-sharing system found that trip duration increased by 88% from an average of 23 minutes to 43 minutes per trip. They also found that due to this disruption the bicycle trip counts increased by 85% from an average of 38,886 trips per day to 72,503 trips per day [16]. Fuller et al. [32] investigating the effect of London Tube strikes on 6 September and 10 October 2010 on the London bike-sharing system, found that these strikes did not cause any significant increase in mean trip duration. However, a statistically significant increase in the total number of cycle hires per day was observed [32]. Both studies concluded that changes in the system caused by the above-mentioned disruptions were temporary. Similarly,

Younes et al. [17] investigated the impact of three planned disruptive events in Washington D.C. metro services on Capital bikeshare. They found that, while disruptions had increased bike ridership significantly, the change in the mean hire duration was insignificant because the increase in the hire numbers for trips longer than 2.5 miles were relatively small.

Previous studies on understanding the effect of the outbreak of communicable diseases on bike-sharing schemes are relatively rare [12, 33]. A recent study conducted by Wang and Noland [12] examined the effect of the lockdown and the subsequent phases of reopening on Citi Bike in New York, analysing two years of data, 2019 and 2020. The authors used a Prais-Winsten regression model that accommodates serial correlation given the time-series nature of their data. They found that the demand for the Citi Bike system decreased sharply after the lockdown, but it started to return normal afterwards. Another recent study conducted by Li et al. [34] found that travellers, in Zurich, preferred to use micro-mobility services (including bike-sharing services) during the COVID-19 pandemic, and that these services were used for longer trips. Similarly, Li et al. [35] analysing the demand for the London bike-sharing system over the period January 2019 to June 2020, estimated the effect of the first lockdown and lockdown ease on the number of daily trips. They found that the number of trips decreased after the lockdown, but then the demand showed an increasing trend.

## 1.2 The current paper

In this research we investigate the effect of the COVID-19 pandemic on the London bike-sharing scheme. While previous research in this context is limited and mostly focuses on understanding the effect on the number of trips, we estimate the impact of the pandemic on both hire time (trip duration) and hire numbers in London, UK. Also, while previous studies provide valuable insights, they mostly use pandemic-related policy interventions as explanatory variables in regression models to estimate the effect of these interventions on bike-sharing systems based on relatively short time spans (one or two years of data). In this study, however, considering time-series data from 2010 to 2020, we use a Bayesian second-order random walk time-series model to predict what would have happened if the COVID-19 pandemic had not happened; that is, the counterfactual. The model accounts for time dependency in our time-series data as well as the non-linear effect of time on the outcomes of interest: hire time and hire numbers. We then compare the observed hire numbers and hire time (trip duration) with their respective counterfactuals to estimate the varying effect of the COVID-19 pandemic over the period March-December 2020. This allows us to understand the varying effect of the pandemic on hire numbers and hire time during the latter period as various pandemic-related policies came into force. Our method is one of the most valid approaches used in biostatistics and medical research; for example, to estimate excess mortality during the recent pandemic [36-37].

## 2. Data

The data set used in this study is related to the London bike-sharing system and was obtained from Transport for London (TfL). The data set contains average monthly hire time (trip duration) and total monthly cycle hire numbers over an 11-year period from the introduction of the scheme in July 2010 to December 2020. Note that the data used here — which was readily available on TfL's website — is at an aggregate level; i.e., for the entire London bike-sharing system at a monthly level. The study period therefore covers 126 months in total. To control for the size of the system, we obtained time-series of the number of docking stations from TfL, aggregated at a monthly level. Also, weather-related variables (rainfall, temperature, humidity, and wind) were obtained from UK Met Office Integrated Data Archive System and NW3 weather website, and were aggregated at a monthly level to be in accordance with the aggregation level of the outcomes of interest (hire duration and hire numbers). Table 1 provides the summary statistics of the data. Fig. 1 displays time-series of the outcomes. This figure implies no major change in the pattern of the cycle hires in 2020 compared to the previous years. However, we see that the pattern changes drastically for hire time in 2020.

Table 1. Summary of descriptive statistics (July 2010 – December 2020)

| Variable | Mean | Std. Dev. | Min | Max |
|---|---|---|---|---|
| Average monthly hire time (minutes) | 19.28 | 3.63 | 13.78 | 36.00 |
| Monthly number of cycle hires | 785,366.00 | 237,647.80 | 12,461.00 | 1,253,102.00 |
| Monthly number of docking stations | 687.56 | 158.47 | 315.00 | 834.00 |
| Temperature (°C) | 12.14 | 4.85 | 2.01 | 22.11 |
| Rainfall (mm) | 1.73 | 0.99 | 0.13 | 5.37 |
| Wind (mph) | 4.90 | 1.01 | 2.77 | 8.67 |
| Humidity (%) | 75.54 | 8.12 | 60.06 | 90.33 |

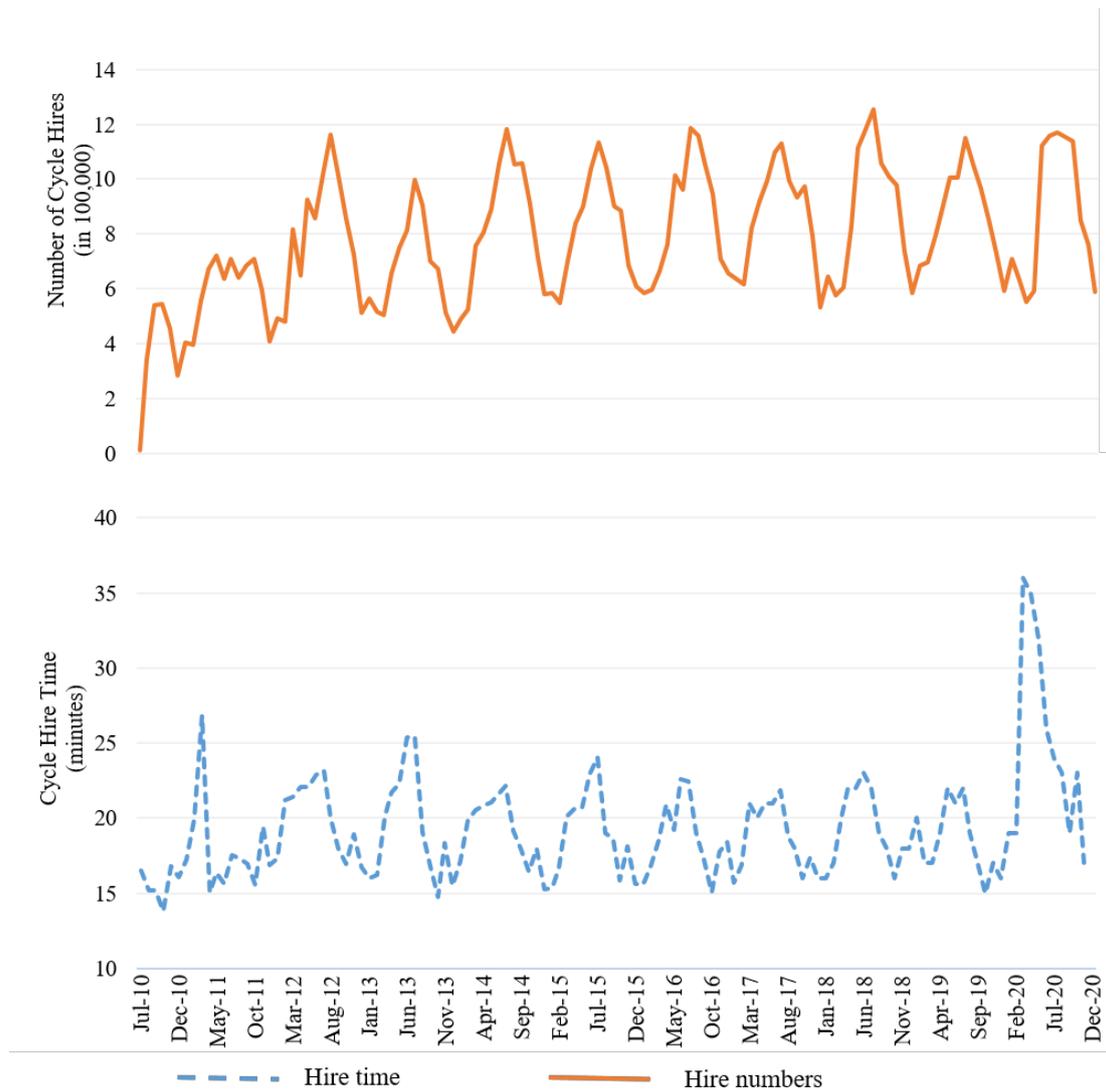

Figure 1. London bike-share time-series of monthly hire numbers and average monthly hire time

Fig. 2 takes a closer look into the year 2020, displaying the observed trend in hire numbers and hire time as well as some of the major events (policies) relating to the COVID-19 pandemic. These events are obtained from the Institute for Government Analysis [38]. The first lockdown in England was implemented on 26th March 2020, and since then several changes to restrictions were made by the Government. In May, people who could not work from home were told to go to work, avoiding public transit. Other similar changes to the pandemic-related policies were made over the period May-December 2020. Perhaps, the most important one being the second lockdown that came into force on 5 November 2020, ending on 2 December 2020. It can be seen that the outcomes of interest follow two different trends. Fluctuations in the graph are partly due to seasonal effects and partly due to the implemented policies. It is therefore important to distinguish between the two major sources that influence the London bike-sharing system — which we will discuss in the next section.

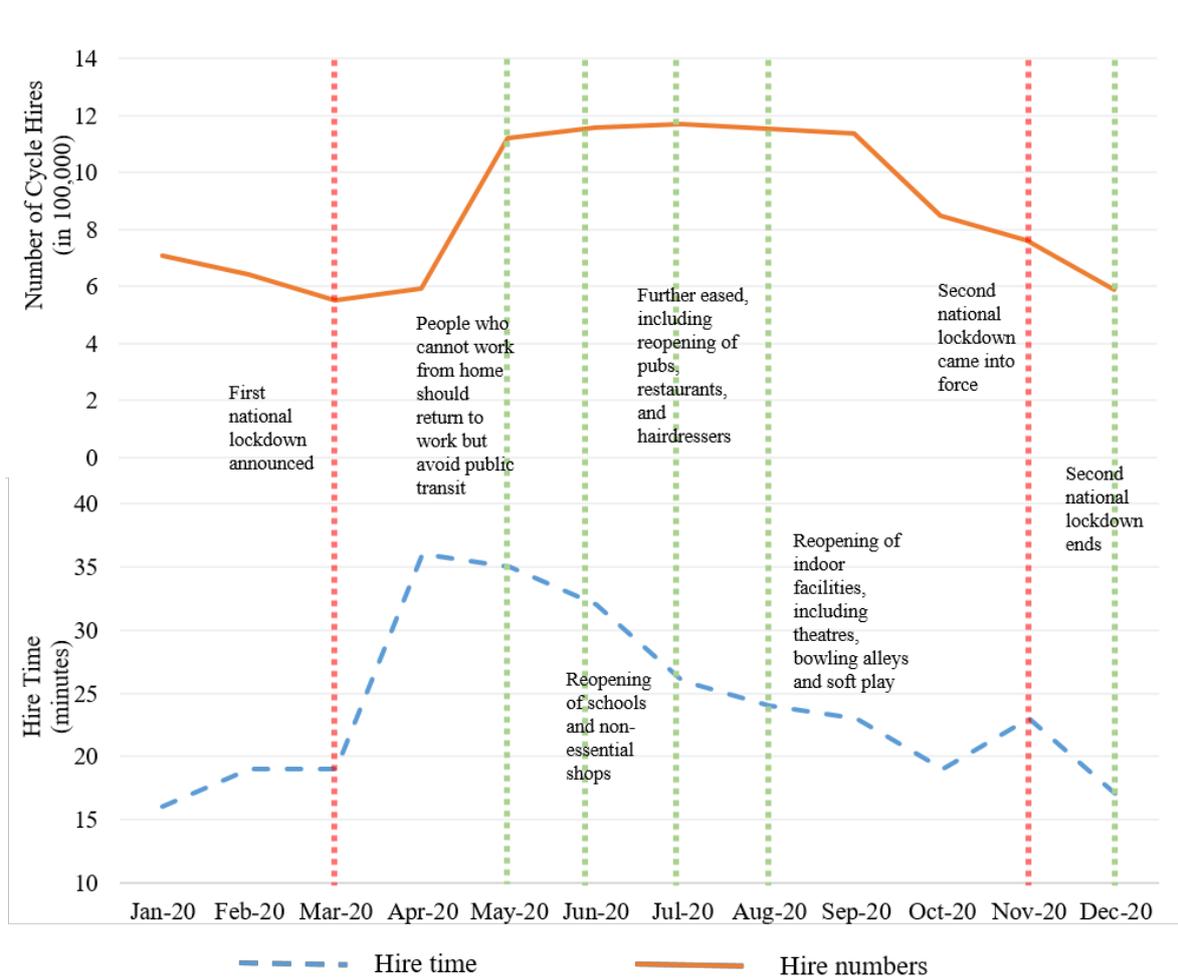

Figure 2. Time-series of observed data in 2020 and pandemic-related events

## 3. Methodology

To estimate the change in the London bike-sharing system in terms of both hire number and hire time, we used the pre-lockdown period data from July 2010 to the end of February 2020 (training data) to calibrate our statistical models. We used Bayesian hierarchical models with a second-order random walk specification to account for the temporal correlation in our time-series data. We then used the data from March 2020 to December 2020 to predict hire numbers and hire time for each month had the pandemic not occurred; that is, counterfactuals. To help predictions, we considered a set of covariates including the meteorological variables, number of docking stations, and different lagged versions of the outcome (1, 2, 6 , and 12-month lag) based on the previous literature and association with the outcomes of interest (based on parsimony grounds). Finally, comparing the counterfactuals with the observed data in the post-lockdown period, we were able to understand how the recent pandemic affected the London bike-sharing system March-December 2020. A schematic view of the method is displayed in Fig. 3.

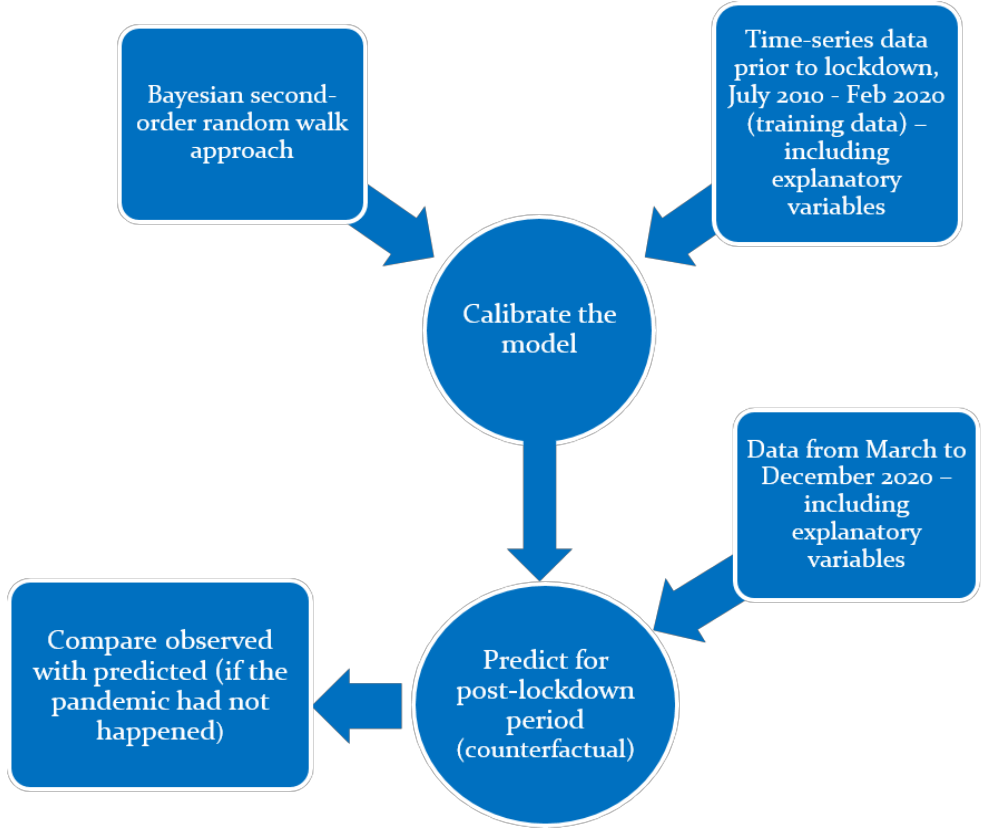

Figure 3. Schematic view of the methodological approach

## 4.1. Bayesian hierarchical second-order random walk model

Hire time is a continuous variable while the monthly cycle hire numbers are large counts. To make the monthly cycle hire numbers continuous and help the convergence, we standardised this variable, subtracting the mean and dividing by the standard deviation, and back-transformed the predictions. Let $y_t$ denote the observed outcome of interest (e.g. hire time or standardized hire numbers) for the $t$-th month ($t = 1, 2, …, T$). It is assumed that $y_t$ follows a normal density with the mean $\lambda_t$ and variance $v$. Let $X = (X_1, X_2,…, X_k)$ be the vector of explanatory variables (e.g., weather conditions) with their corresponding regression parameters $\beta = (\beta_1, \beta_2,…, \beta_k)$, and $\beta_0$ denotes an intercept term. Let also $\gamma$ be the coefficient associated with the 12-month lag of the outcome. We can write

$$y_t \sim Normal(\lambda_t, v)$$
$$\lambda_t = \beta_0 + X_t.\beta + \gamma.y_{t-12} + u_t \tag{1}$$

where $u_t$ — specified in (2) — is a structured error term that follows a second-order random walk (RW2) process, accommodating temporal correlation in the time-series data. Note that $u_t$ allows for non-linearity in the effect of time on the outcome of interest.

$$p(u_t|u_{-t}, v_e) = \begin{cases} Normal(2u_{t+1} - u_{t+2}, v_e) & \text{for t=1} \\ Normal(\frac{2}{5}u_{t-1} + \frac{4}{5}u_{t+1} - \frac{1}{5}u_{t+2}, v_e/5) & \text{for t=2} \\ Normal\left(-\frac{1}{6}u_{t-2} + \frac{2}{3}u_{t-1} + \frac{2}{3}u_{t+1} - \frac{1}{6}u_{t+2}, v_e/6\right) & \text{for t=3,...,T-2} \quad (2) \\ Normal\left(-\frac{1}{5}u_{t-2} + \frac{4}{5}u_{t-1} + \frac{2}{5}u_{t+1}, v_e/5\right) & \text{for t=T-1} \\ Normal(-u_{t-2} + 2u_{t-1}, v_e) & \text{for t=T} \end{cases}$$

We specified non-informative priors Normal(0, 1000) for the regression coefficients, and Gamma densities with shape 1 and rate 0.01 for $v$ and $v_e$. The rationale behind this selection is to have an adequate mass at zero, making sure that a more complex model is not forced to the data, but driven by the data. We report median and 95% credible intervals of the model predictions and the regression coefficients. The covariates were standardised to help convergence. We estimated the models using Nimble (https://r-nimble.org/).

### 4.2. Leave-one-year out cross-validation

To investigate model performance, we employed a comprehensive cross-validation approach. In this approach, we focused on the years 2010-2019 and fit the aforementioned models leaving one year out each time. We then predicted hire number and hire time for the year left out and compared the predicted with the observed values. As metrics, we calculated the adjusted $R^2$ and the 95% coverage probability, which is the probability that an observed value lies within the 95% credible intervals of the predictions.

### 4. Results and discussion

Following our cross-validation exercise, the adjusted $R^2$ values were 0.69 and 0.30 for the models representing cycle hire number and hire time, respectively. These values are satisfactory when comparing with previous research. For example, the adjusted $R^2$ value for trip duration was less than 0.19 in an important study conducted by Gebhart and Noland [24]. The 95% coverage probabilities were 0.86 and 0.94 for the hire number and hire time models, respectively. The coverage probability indicates the proportion of the observed data that falls within the predicted 95% credible intervals. We are therefore satisfied with the performance of the developed models.

Although the focus of our study is on predicting what would have happened if the COVID-19 pandemic had not occurred, we provide the regression coefficient estimates for both hire time and hire number models in Table 2. This allows identify explanatory variables that are statistically important in explaining monthly cycle hires and hire time. Note that, as mentioned in Section 4, explanatory variables were standardised so that the interpretation is: changes in the outcome for every standard deviation increase in the covariates. In accordance with previous research, we found that weather-related variables have an important effect on the demand and trip duration. While temperature is positively associated with these two measures, rainfall, humidity, and wind are negatively associated with cycle hire numbers and hire time. This is in accordance with previous research. Also, we found that the lag of 12 (i.e., $y_{t-12}$) was an important predictor for both outcomes. This can be explained by the fact that for each month, for example, travel patterns and weather conditions are similar to that month's observations in the previous year.

Table 2. Posterior summary of the regression coefficients

| Variables[2] | Monthly cycle hire numbers[1] | | | Average monthly hire time | | |
|---|---|---|---|---|---|---|
| | | 95% credible intervals | | | 95% credible intervals | |
| | Median | lower limit | upper limit | Median | lower limit | upper limit |
| Temperature | 0.71 | 0.61 | 0.80 | 1.14 | 0.79 | 1.48 |
| Rainfall | -0.14 | -0.19 | -0.09 | - | - | - |
| Wind | -0.09 | -0.14 | -0.03 | - | - | - |
| Humidity | - | - | - | -0.72 | -1.10 | -0.33 |
| Lag 12 | 0.18 | 0.06 | 0.29 | 0.66 | 0.29 | 1.03 |

1 Hire numbers are standardised

2 Explanatory variables are standardised

## 5.1. Observed post-lockdown data vs. counterfactuals

Fig. 4 displays the observed and predicted (counterfactual) trends for cycle hire numbers in 2020. Specifically, the observed and predicted cycle hire numbers with their 95% credible intervals (the shaded area) are shown in Fig. 4 over the period January-December 2020. Recall that the first lockdown came into force in March 2020 in England. As it can be seen in Fig. 4, the number of cycle hires decreased in March and April, which is expected following the city shut down as of March 23rd. It is interesting that the demand in March experienced a relatively important decrease although the lockdown started on 23 March 2020. This could be explained partly due to the lockdown and partly because of the fact that some travellers started to work from home (or reduced their number of trips) in March 2020 even prior to the UK Government's lockdown policy comes into force.

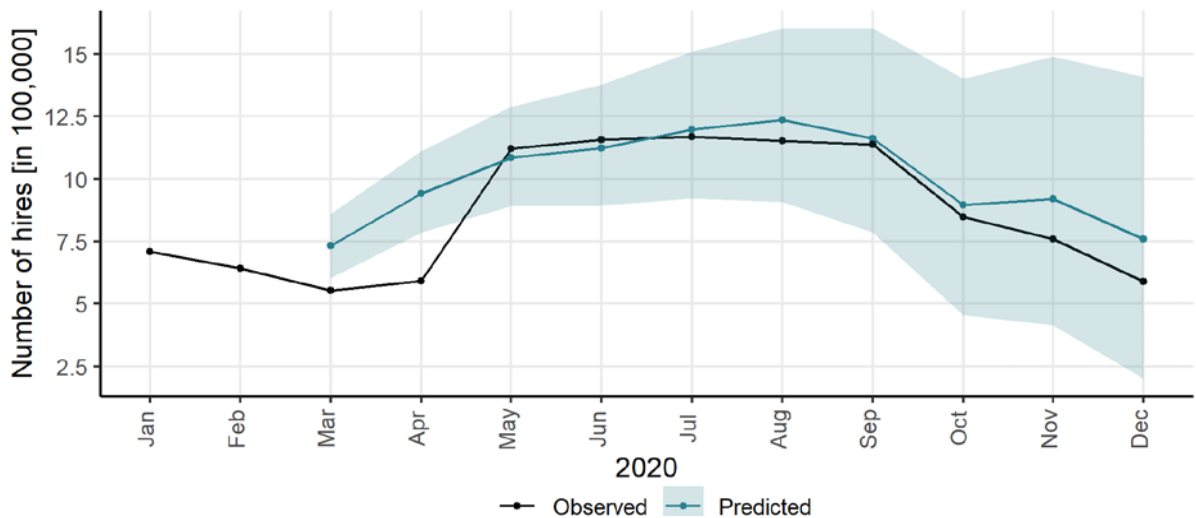

Figure 4. Observed hire numbers vs. predicted hire numbers (counterfactuals). Note: the shaded area indicates the 95% credible intervals around counterfactuals.

After this reduction, the demand rebounded from May 2020, and remained in the expected range of what would have been if the pandemic had not occurred. This could be an indication that the London bike-sharing scheme has been a resilient transport system during the year 2020 in spite of the pandemic. Previous studies highlighted the resiliency of bikesharing, for example, in New York [12, 33]. In May 2020 some of the restrictions were eased; for example, people who could not work from home were told to go to work, avoiding public transport if they can. Therefore, while public transit suffered in terms of ridership [39, 40], we see a slight increase in the number of cycle hires in May and June 2020. In the period May-December 2020, the larger decrease in the demand was observed in November and December after the second lockdown came into force on 5th November.

Fig. 5 shows the observed and predicted average monthly hire time with their associated 95% credible intervals. The pattern differs from the one in Fig. 4. In March while the average hire time increased slightly following the first lockdown on 23rd March, an important increase occurred in April, May, and June. Then, from July 2020, the average hire time remained within the posterior distribution of the predicted hire time; therefore, no statistically distinguishable change was observed. Interestingly, in October 2020 hire time became very similar to what it would have been if the pandemic had not happened. Then, following the second lockdown introduced during the first week of November, there was another jump in hire time. The second lockdown ended on 2nd December, and the average hire time in December became similar to its corresponding counterfactual.

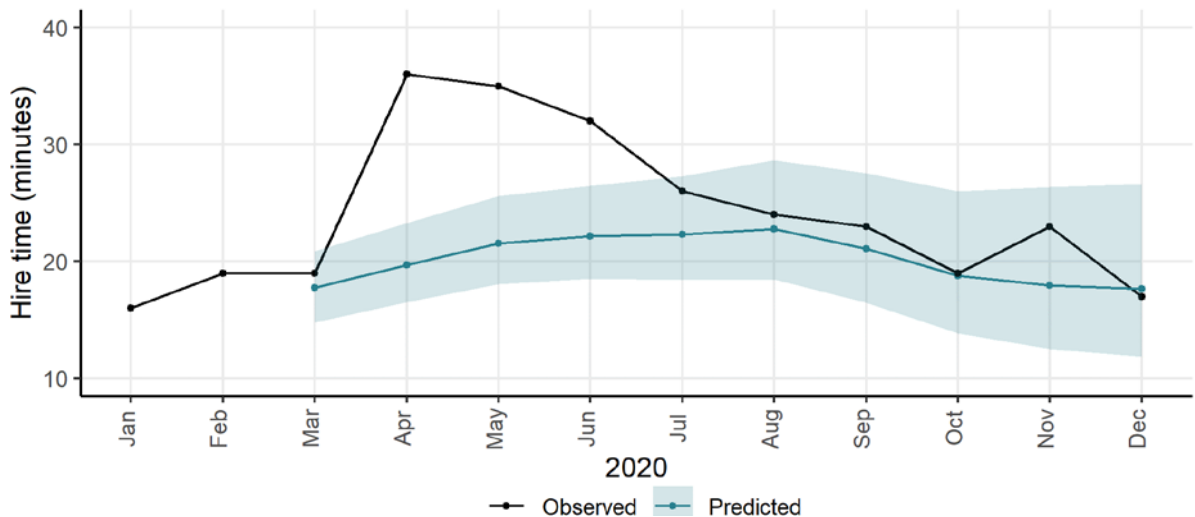

Figure 5. Observed hire time vs. predicted hire time (counterfactuals). Note: the shaded area indicates the 95% credible intervals around counterfactuals.

The increase in hire time could be partly associated with the fact that perhaps some travellers hired bikes for longer trips, avoiding tube and buses in London. As indicated by previous research, there has been a shift from public transit to other modes of travel including bikeshare; for instance, in New York due to the fear of contacting the Coronavirus while travelling on public transport [12, 33]. Another reason for the observed increase in hire time could be partly associated with an increase in causal users. For example, this might be due to an increase in the number of Londoners who used Santander Cycles for recreational activities including exercising. On the other hand, note that according to official statistics, the number of tourists were lower compared to the previous years over the same period [41].

### 5.2. Magnitude of the change in hire numbers and hire time

To clearly understand the magnitude of the change in hire numbers and hire time during 2020, we obtained posterior densities of the estimated changes (See Table 3). With respect to the total number of monthly cycle hires, Table 3 indicates that the largest decrease occurred in April with around 360,000 fewer bikes being hired, followed by March with around 184,000 fewer bikes being hired. Looking at the 95% credible intervals, the reductions in March and April 2020 are statistically important.

Table 3. Estimated change in the London bike-sharing scheme

| Month 2020 | Monthly cycle hires (numbers) 95% credible intervals | | | Monthly average cycle hire time (minutes) 95% credible intervals | | |
|---|---|---|---|---|---|---|
| | Median | Lower limit | Upper limit | Median | Lower limit | Upper limit |
| Mar | *-183,849* | *-311,416* | *-57,143* | 1.39 | -1.72 | 4.44 |
| Apr | *-359,531* | *-525,787* | *-202,016* | *16.48* | *12.97* | *19.8* |
| May | 25,377 | -179,939 | 220,883 | *13.72* | *9.88* | *17.42* |
| Jun | 21,602 | -232,512 | 254,274 | *10.11* | *5.86* | *13.95* |
| Jul | -46,989 | -350,991 | 233,979 | 3.96 | -0.72 | 8.25 |
| Aug | -106,165 | -464,076 | 227,011 | 1.52 | -3.87 | 6.31 |
| Sept | -50,532 | -480,584 | 333,468 | 2.24 | -3.67 | 7.45 |
| Oct | -74,784 | -567,513 | 366,869 | 0.6 | -5.96 | 6.28 |
| Nov | -191,748 | -743,497 | 311,631 | 5.47 | -1.98 | 11.75 |
| Dec | -204,757 | -830,529 | 363,241 | -0.19 | -8.23 | 6.43 |

With respect to hire time, the largest statistically important change in trip duration occurred in April with an excess of 16.48 minutes (CI[12.97, 19.18]), followed by May with an excess of 13.72 (CI[9.88, 17.42]) minutes, and then June with an excess of 10.11 (CI[5.86, 13.95]) minutes. The 95% intervals indicate the level of uncertainty around the estimates. Note that the interpretation is different from classical confidence intervals. A 95% credible interval indicates that there is 95% probability that the estimated value (median) is in that interval. In July the excess hire time decreased to 3.96 (CI[-0.72, 8.25]) minutes. Although the latter interval includes zero, a larger proportion of the posterior distribution is on the positive side. The same trend is observed in November 2020 when the second national lockdown came into force.

### 5.3. Strengths and limitations of the study

Our method allowed us to understand the varying effect of the COVID-19 pandemic on the London bike-sharing system over the period March-December 2020, using a rigorous time-series model. Specifically, based on our approach, we were able to investigate how the London bike-sharing scheme was affected by various Government policies (i.e., introduction of the lockdown and the easing of restrictions in various stages) that came into force in relation to the COVID-19 pandemic. The Bayesian approach is advantageous as it allows for the propagation of uncertainties in all layers of the model and predictions. Also, the predictions under the second-order random walk approach are smoother (using more neighbouring time periods), which makes it appealing in the context of our study in which the focus is on prediction [37]. Lastly, the leave-one-year-out cross-validation approach adopted here allowed us to ensure the suitability of the models and predictions. In general, using such cross-validation approach is rare if non-existent in the bikesharing literature. Also, while previous research mostly focused on a relatively limited time span in studying bikesharing systems, we considered an eleven-year study period, from the launch of the London bike-sharing scheme to the end of December 2020.

To provide more evidence and better understand the magnitude of the shift from public transport to bikesharing, analysing public transit data together with the bikeshare data is needed. Also, the data used in this study was at an aggregate level in terms of both user type and docking station level. Firstly, analysing the data relating to subscriber and casual users (two different segments of the users) helps understand the impact of the recent pandemic on the London bike-sharing system more fully. Secondly, another important improvement will be achieved by investigating the change at station level. Doing so, it is possible to understand how the effect of the pandemic varies from one docking station to another.

For instance, we expect that docking stations in proximity to offices where most users are employees, working mostly from home, are more affected by the pandemic compared to the docking stations located in commercial or residential areas.

## 5. Conclusions and implications

The aim of our study was to investigate the impact of the COVID-19 pandemic on the London bike-sharing system, which is one of the largest bike-sharing schemes in the world. To this end, we focused on the readily available data available on TfL's website. The data contained the total number of monthly cycle hires and the average monthly cycle hire time (trip duration) from July 2010 (launch of the London bike-sharing scheme) to the end of December 2020. In addition, we obtained weather-related data and time-series of docking stations from various sources. Using a Bayesian second-order random walk time-series approach, we calibrated a model using the pre-lockdown data (July 2010-February 2020). Then, we predicted what would have been if the pandemic had not occurred; that is, counterfactuals for both hire numbers and hire time for the post-lockdown period from March to December 2020. Comparing the observed data during the post-lockdown period with the counterfactuals and their associated 95% credible intervals, we examined how the London bike-sharing system was affect by the pandemic. Statistically distinguishable changes occurred in March and April with respect to hire numbers, and in April, May, and June with respect to hire time.

The interactions of travellers and travel mode characteristics in London would give rise to how the Government's COVID-related policies has affected the London bike-sharing system. The fact that some travellers might have shifted from public transit to bikeshare has important implications as this could result in a permanent (or at least relatively long-term) change in their travel behaviour, a success that could have not been achieved easily if the pandemic had not happened or if the pandemic-related policies had not been introduced. A discussion on the possibility that this behavioural change may become permanent is provided by Wang and Noland [12]. To reveal the underlying mechanisms behind such behavioural changes, conducting travel surveys would provide valuable insights.

Note that a part of the shift from public transport may have been towards driving alone, especially where active travel is less favourable. Therefore, our recommendation is to extend bikesharing in a way that it covers the entire Greater London area. The fact that the number of cycle hires has not experienced any statistically important reduction from May to December 2020 would indicate that the London bike-sharing scheme has been a resilient transportation system during the pandemic. Therefore, it is important to integrate micro mobility (e.g., bike-sharing schemes) in urban transportation systems not only to increase their resiliency but also to improve air quality; and consequently, human health. At the same time, promoting active modes of travel, cycling and walking should become a priority in urban areas.

**Acknowledgements**

Garyfallos Konstantinoudis is supported by an MRC Skills Development Fellowship [MR/T025352/1]. We would like to thank Transport for London for providing the data.

**Conflicts of interest**

The authors declare no conflict of interest.

# References

1. Pucher J, Buehler R. Making Cycling Irresistible: lessons from the Netherlands, Denmark and Germany. Transport Reviews. 2008; 28:495–528. doi:10.1080/01441640701806612.
2. Shaheen SA, Guzman S, Zhang H. Bikesharing in Europe, the Americas, and Asia. Transportation Research Record. 2010; 2143:159–67. doi:10.3141/2143-20.
3. Pucher J, Dill J, Handy S. Infrastructure, programs, and policies to increase bicycling: an international review. Preventive Medicine. 2010; 50 Suppl 1:S106-25. doi:10.1016/j.ypmed.2009.07.028.
4. Caggiani L, Camporeale R, Binetti M, Ottomanelli M. An urban bikeway network design model for inclusive and equitable transport policies. Transportation Research Procedia. 2019; 37:59–66. doi:10.1016/j.trpro.2018.12.166.
5. Li H, Zhang Y, Ding H, Ren G. Effects of dockless bike-sharing systems on the usage of the London Cycle Hire. Transportation Research Part A: Policy and Practice. 2019; 130:398–411. doi:10.1016/j.tra.2019.09.050.
6. Cervero R, Denman S, Jin Y. Network design, built and natural environments, and bicycle commuting: evidence from British cities and towns. Transport Policy. 2019; 74:153–64. doi:10.1016/j.tranpol.2018.09.007.
7. Eren E, Uz VE. A review on bike-sharing: the factors affecting bike-sharing demand. Sustainable Cities and Society. 2020; 54:101882. doi:10.1016/j.scs.2019.101882.
8. Chibwe J, Heydari S, Faghih Imani A, Scurtu A. An exploratory analysis of the trend in the demand for the London bike-sharing system: from London Olympics to Covid-19 pandemic. Sustainable Cities and Society. 2021; 69:102871. doi:10.1016/j.scs.2021.102871.
9. Transport for London. TfL and Santander celebrate ten years of London's flagship cycle hire scheme. 2020. https://tfl.gov.uk/info-for/media/press-releases/2020/january/tfl-and-santander-celebrate-ten-years-of-london-s-flagship-cycle-hire-sche. Accessed on 5/21/2021.
10. Borkowski P, Jażdżewska-Gutta M, Szmelter-Jarosz A. Lockdowned: everyday mobility changes in response to COVID-19. Journal of Transport Geography. 2021; 90:102906. doi:10.1016/j.jtrangeo.2020.102906.
11. Wang K-Y. How change of public transportation usage reveals fear of the SARS virus in a city. PLoS One. 2014; 9:e89405. doi:10.1371/journal.pone.0089405.
12. Wang H, Noland RB. Bikeshare and subway ridership changes during the COVID-19 pandemic in New York City. Transport Policy. 2021; 106:262–70. doi:10.1016/j.tranpol.2021.04.004.
13. Nikitas A, Tsigdinos S, Karolemeas C, Kourmpa E, Bakogiannis E. Cycling in the Era of COVID-19: Lessons Learnt and Best Practice Policy Recommendations for a More Bike-Centric Future. Sustainability. 2021; 13:4620. doi:10.3390/su13094620.
14. Noland RB. Mobility and the effective reproduction rate of COVID-19. Journal of Transport & Health. 2021; 20:101016. doi:10.1016/j.jth.2021.101016.
15. Mattsson L-G, Jenelius E. Vulnerability and resilience of transport systems – A discussion of recent research. Transportation Research Part A: Policy and Practice. 2015;81:16–34. doi:10.1016/j.tra.2015.06.002.
16. Saberi M, Ghamami M, Gu Y, Shojaei MH, Fishman E. Understanding the impacts of a public transit disruption on bicycle sharing mobility patterns: a case of Tube strike in London. Journal of Transport Geography. 2018;66:154–66. doi:10.1016/j.jtrangeo.2017.11.018.
17. Younes H, Nasri A, Baiocchi G, Zhang L. How transit service closures influence bikesharing demand; lessons learned from SafeTrack project in Washington, D.C. metropolitan area. Journal of Transport Geography. 2019; 76:83–92. doi:10.1016/j.jtrangeo.2019.03.004.
18. Martin EW, Shaheen SA. Evaluating public transit modal shift dynamics in response to bikesharing: a tale of two U.S. cities. Journal of Transport Geography. 2014; 41:315–24. doi:10.1016/j.jtrangeo.2014.06.026.

19. Faghih-Imani A, Eluru N, El-Geneidy AM, Rabbat M, Haq U. How land-use and urban form impact bicycle flows: evidence from the bicycle-sharing system (BIXI) in Montreal. Journal of Transport Geography. 2014; 41:306–14. doi:10.1016/j.jtrangeo.2014.01.013.
20. Guo Y, Zhou J, Wu Y, Li Z. Identifying the factors affecting bike-sharing usage and degree of satisfaction in Ningbo, China. PLoS One. 2017; 12:e0185100. doi:10.1371/journal.pone.0185100.
21. Li H, Ding H, Ren G, Xu C. Effects of the London Cycle Superhighways on the usage of the London Cycle Hire. Transportation Research Part A: Policy and Practice. 2018; 111:304–15. doi:10.1016/j.tra.2018.03.020.
22. Nikitas A. How to Save Bike-Sharing: an Evidence-Based Survival Toolkit for Policy-Makers and Mobility Providers. Sustainability. 2019; 11:3206. doi:10.3390/su11113206.
23. Lazarus J, Pourquier JC, Feng F, Hammel H, Shaheen S. Micromobility evolution and expansion: understanding how docked and dockless bikesharing models complement and compete – A case study of San Francisco. Journal of Transport Geography. 2020; 84:102620. doi:10.1016/j.jtrangeo.2019.102620.
24. Gebhart K, Noland RB. The impact of weather conditions on bikeshare trips in Washington, DC. Transportation. 2014; 41:1205–25. doi:10.1007/s11116-014-9540-7.
25. Shaheen SA, Martin E, Chan ND, Cohen AP, Pogodzinski M. Public Bikesharing in North America during a Period of Rapid Expansion: understanding Business Models, Industry Trends, and User Impacts. 2014. Available online at https://transweb.sjsu.edu/research/Public-Bikesharing-North-America-During-Period-Rapid-Expansion-Understanding-Business-Models-Industry-Trends-and-User-Impacts, accessed on 5/21/2021.
26. Faghih-Imani A, Eluru N. A finite mixture modeling approach to examine New York City bicycle sharing system (CitiBike) users' destination preferences. Transportation. 2020; 47:529–53. doi:10.1007/s11116-018-9896-1.
27. Sears J, Flynn BS, Aultman-Hall L, Dana GS. To Bike or Not to Bike. Transportation Research Record. 2012; 2314:105–11. doi:10.3141/2314-14.
28. El-Assi W, Salah Mahmoud M, Nurul Habib K. Effects of built environment and weather on bike sharing demand: a station level analysis of commercial bike sharing in Toronto. Transportation. 2017; 44:589–613. doi:10.1007/s11116-015-9669-z.
29. Morton C. The demand for cycle sharing: examining the links between weather conditions, air quality levels, and cycling demand for regular and casual users. Journal of Transport Geography. 2020; 88:102854. doi:10.1016/j.jtrangeo.2020.102854.
30. Kim K. Investigation on the effects of weather and calendar events on bike-sharing according to the trip patterns of bike rentals of stations. Journal of Transport Geography. 2018; 66:309–20. doi:10.1016/j.jtrangeo.2018.01.001.
31. Weinert J, Ma C, Cherry C. The transition to electric bikes in China: history and key reasons for rapid growth. Transportation. 2007; 34:301–18. doi:10.1007/s11116-007-9118-8.
32. Fuller D, Sahlqvist S, Cummins S, Ogilvie D. The impact of public transportation strikes on use of a bicycle share program in London: interrupted time series design. Preventive Medicine. 2012; 54:74–6. doi:10.1016/j.ypmed.2011.09.021.
33. Teixeira JF, Lopes M. The link between bike sharing and subway use during the COVID-19 pandemic: the case-study of New York's Citi Bike. Transportation Research Interdisciplinary Perspective. 2020; 6:100166. doi:10.1016/j.trip.2020.100166.
34. Li A, Zhao P, He H, Axhausen KW. Understanding the variations of micro-mobility behavior before and during COVID-19 pandemic period: ETH Zurich; 2020.
35. Li H, Zhang Y, Zhu M, Ren G. Impacts of COVID-19 on the usage of public bicycle share in London. Transportation Research Part A: Policy and Practice. 2021; 150:140–55. doi:10.1016/j.tra.2021.06.010.
36. Kontis V, Bennett JE, Rashid T, Parks RM, Pearson-Stuttard J, Guillot M, et al. Magnitude, demographics and dynamics of the effect of the first wave of the COVID-19 pandemic on all-

cause mortality in 21 industrialized countries. Nature Medicine. 2020; 26:1919–28. doi:10.1038/s41591-020-1112-0.
37. Blangiardo M, Cameletti M, Pirani M, Corsetti G, Battaglini M, Baio G. Estimating weekly excess mortality at sub-national level in Italy during the COVID-19 pandemic. PLoS One. 2020; 15:e0240286. doi:10.1371/journal.pone.0240286.
38. Haddon C, Sasse T, Tetlow G. Lifting lockdown in 2021: the next phase of the government's coronavirus strategy. 2021. https://www.instituteforgovernment.org.uk/publications/lifting-lockdown. Accessed on 3/28/2021.
39. Department for Transport. Quarterly Bus Statistics: England, April to June 2020. 2020. https://www.gov.uk/government/statistics/quarterly-bus-statistics-april-to-june-2020. Accessed on 6/1/2021.
40. Vickerman R. Will Covid-19 put the public back in public transport? A UK perspective. Transport Policy. 2021; 103:95–102. doi:10.1016/j.tranpol.2021.01.005.
41. Office of National Statistics. Coronavirus and the impact on the UK travel and tourism industry. 2021. https://www.ons.gov.uk/businessindustryandtrade/tourismindustry/articles/coronavirusandtheimpactontheuktravelandtourismindustry/2021-02-15. Accessed 6/1/2021.